\documentstyle[epsfig,aps,preprint]{revtex} 
\begin{document}
\input DraTex.sty         

\def\d{\delta}
\def\bea{\begin{eqnarray}}
\def\eea{\end{eqnarray}}
\def\beas{\begin{eqnarray*}}
\def\eeas{\end{eqnarray*}}
\def\nn{\nonumber}
\def\ni{\noindent}

\def\G{\Gamma}
\def\S{\Sigma}
\def\Y{\Psi}
\def\Yb{\bar\Psi}
\def\F{\Phi}
\def\W{\Omega}
\def\d{\delta}
\def\l{\lambda}
\def\r{\rho}
\def\g{\gamma}
\def\m{\mu}
\def\n{\nu}
\def\s{\sigma}
\def\tt{\theta}
\def\b{\beta}
\def\a{\alpha}
\def\k{\kappa}
\def\t{\tau}
\def\f{\phi}
\def\fh{\hat{\phi}}
\def\y{\psi}
\def\z{\zeta}
\def\p{\pi}
\def\h{\eta}
\def\e{\epsilon}
\def\ve{\varepsilon}
\def\vf{\varphi}
\def\fc{\phi_{\rm c}}
\def\ck{{\cal K}}
\def\cl{{\cal L}}
\def\cd{{\cal D}}
\def\cm{{\cal M}}
\def\cv{{\cal V}}
\def\ch{{\cal H}}
\def\cz{{\cal Z}}
\def\pl{\partial}
\def\ov{\over}
\def\~{\tilde}
\def\rar{\rightarrow}
\def\lar{\leftarrow}
\def\lrar{\leftrightarrow}
\def\rra{\longrightarrow}
\def\lla{\longleftarrow}
\def\ot{\otimes}
\def\<{\langle}
\def\>{\rangle}
\def\8{\infty}

\def\la{\LineAt}
\def\lt{\LineTo}
\def\li{\Line}
\def\mt{\MoveTo}
\def\mo{\Move}
\def\ml{\MarkLoc}
\def\mtl{\MoveToLoc}
\def\ltl{\LineToLoc}
\def\dra{\DrawRectAt}
\def\dr{\DrawRect}
\def\dc{\DrawCircle}
\def\do{\DrawOval}
\def\doa{\DrawOvalArc}
\def\pra{\PaintRectAt}
\def\pr{\PaintRect}
\def\pc{\PaintCircle}
\def\poa{\PaintOvalArc}
\def\po{\PaintOval}
\def\te{\Text}
\def\un{\Units}
\def\sc{\Scale}
\def\ent{\EntryExit}
\def\ps{\PenSize}
\def\ra{\Ragged}
\def\rta{\RotatedAxes}
\def\era{\EndRotatedAxes}
\def\cn{\CircleNode}


\title{Three-Loop Effective Potential of $O(N)$ $\f^4$ Theory}  
\author{J.~-M. Chung\footnote      
  {Electronic address: jmchung@photon.kyunghee.ac.kr}       
  ~and B.~K. Chung}   
\address{ Research Institute for Basic Sciences\\       
and Department of Physics,\\ Kyung Hee University, Seoul  130-701, Korea}  
  
\date{\today}  
\maketitle  
\draft  
\begin{abstract}       
The three-loop effective potential of the massless $O(N)$ $\f^4$ theory   
is calculated analytically using techniques of dimensional regularization.   
We see a complete agreement between our result and    
Jackiw's result which was obtained only up to two-loop order using a different    
regularization (cutoff regularization) method, but the same renormalization    
conditions. For an easy check of the mutual cancellation of all the dangerous    
pole terms in each loop order, we give the $\e$-expanded loop integrals in    
full detail.       
\end{abstract} 
\section{Introduction}   
The effective potential plays a crucial role in determining the nature of 
the vacuum in a weakly coupled field theory, as was emphasized in the classic
paper of Coleman and Weinberg \cite{cw}. Calculation of this object
by summing infinite series of Feynman diagrams at zero momentum is 
an onerous task, especially when several interactions are present which 
complicate the combinatorial factors that multiply each diagram. 
Jackiw has succeeded in representing each loop order containing
an infinite set of conventional Feynman diagrams by finite number of diagrams
using his algebraic method which can be formally extended to the arbitrary
higher loop order \cite{jk}. 
 
Self-interacting scalar field theory of $\f^4$ model is one of the best
analyzed field theories. Various renormalization group functions of
$O(N)$ $\f^4$ theory are now available up to the five-loop order \cite{sf}.
However, the three-loop effective potential of the same theory is less
progressive. For renormalization of the two-point one-particle-irreducible 
(1PI) Green's function, $\G^{(2)}$, and the four-point 1PI Green's function, 
$\G^{(4)}$, of the $O(N)$ $\f^4$ theory only one type of propagator
is involved in loop integrations, but for calculation of the 
effective potential two kinds of propagators due to two different induced 
mass-like terms are involved in the loop integrations.  
The two-loop effective potential for the massive $O(N)$ $\f^4$ theory in     
four dimensions has been calculated by Ford and Jones \cite{fj}, and for the   
massless $O(N)$ $\f^4$ theory by Jackiw \cite{jk}. For the single-component   
massive $\f^4$ theory in four dimensions, the two-loop effective potential   
has been calculated in Ref.~\cite{iim}, and the three-loop effective potential   
in Ref.~\cite{jm}.   
   
Recently all the genuine three-loop integrals --- genuine in the sense that       
they cannot be factorized into lower-loop integrals --- appearing in the 
three-loop effective potential calculation of the massless $O(N)$ $\f^4$ 
theory has been calculated \cite{jmjm}. 
These integrals carry two kinds of propagator 
lines, A-type line and B-type line. The mass parameter of the B-type line is       
one-third as large as that of the A-type line. 

The purpose of this paper is to calculate the three-loop effective potential
of the massless $O(N)$ $\f^4$ theory in four dimensions of space-time 
in the dimensional regularization scheme \cite{tv}.
In Sec.~II the standard procedeure for the renormalization is presented.   
The Feynman rules for the effective potential suitable for the Jackiw's   
prescription are given and all the Feynman diagrams for the effective  
potential up to three-loop order are constructed. The values of all relevant
integrals are listed in the Appendices A and B. Sec.~III is devoted to  
concluding remarks.

\section{Renormalization of the Effective Potential}       
The Lagrangian for a theory of $N$ spinless fields $\f_a$, with an      
O($N$)-invariant interaction is given as       
\bea       
{\cal L}(\f_a(x))&=&{1+\d Z\ov 2}\pl_\m\f_a\pl^\m\f_a 
-{m^2+\d m^2\ov 2}\f^2-{\l+\d\l\ov 4!}\f^4\;, \label{lg}       
\eea     
where the quadratic and quartic       
expressions are $\f^2=\f_a\f_a$ and      
$\f^4=(\f^2)^2$.     
The quantities $\f_a$, $m$,      
and $\l$ are the renormalized field, the renormalized mass, and  
the renormalized coupling constant respectively, whereas $\d Z$, $\d m^2$,  
and $\d\l$ are corresponding (infinite) counterterm constants.  
These are related to the usual bare quantities as follows:
\beas
m_0^2={m^2+\d m^2\ov 1+\d Z}\;,~~~\l_0={\l+\d \ov (1+\d Z)^2}\;,~~~
\f_0^2=(1+\d Z)\f^2\;.
\eeas
We will confine ourselves     
to the massless theory ($m=0$).  
The effective potential is most suitbly defined generically, when the    
effective action $(\G[\f_{\rm cl}])$, being the generating functional of  
the one-particle-irreducible (1PI) Green's functions  
($\G^{(n)}(x_1,...,x_n)$),   
is expressed in the following local form (the so-called derivative    
expansion):   
\bea   
\G[\f_{\rm cl}]&=&\int\!d^4x\biggl[-\cv(\f_{\rm cl}(x)) 
+{1\ov 2}\cz(\f_{\rm cl}(x))\pl_\m\f_{\rm cl}(x)\pl^\m\f_{\rm    
cl}(x)+\cdots~\biggr]\;,  \label{loc}   
\eea   
where $\f_{\rm cl}(x)$ is the vacuum expectation    
value of the field operator $\f(x)$ in the presence of an external source.    
By setting $\f_{\rm cl}(x)$ in $\cv(\f_{\rm cl}(x))$ to be a constant    
field $\fh$, we obtain the effective potential $V_{\rm eff}(\fh)$   
\bea   
V_{\rm eff}(\fh)\equiv \cv(\f_{\rm cl}(x))|_{\f_{\rm cl}(x)=\fh}\;.\label{ve}   
\eea   
   
Following the field-shift method of Jackiw \cite{jk}   
for the calculation of the effective potential, we first obtain the shifted     
Lagrangian with the constant field configuration $\{\fh_a\}$     
\bea       
{\cal L}(\fh_a;\f_a(x))&=&{1+\d Z\ov 2}\pl_\m\f_a\pl^\m\f_a 
-{1\ov 2}\f_a\biggl[\biggl(\d m^2+{\l+\d\l\ov 6}\fh^2\biggr)\d_{ab}     
+{\l+\d\l\ov 3}\fh_a\fh_b\biggr]\f_b\nn\\     
&-&{\l+\d\l\ov 6}\fh_a\f_a\f^2-{\l+\d\l\ov 4!}\f^4\;, \label{slg}       
\eea     
where $\fh^2=\fh_a\fh_a$.  
Our perturbation theory differs from the bare or renormalized
perturbation theory \cite{ps}; rather a mixed one between both of them. 
The vertices for the $\f$-quadratic term 
with strength $\d Z$ and $\d m^2$ in the usual `renormalized'
perturbation theory are transferred to the propagator line in our treatment.
Thus our perturbation theory follows the one used by Kastening \cite{ks} 
in spirit. The Feynman rules of this shifted Lagrangian are      
\bea     
\Draw\ps(0.8mm)\la(3,3,43,3)\mt(23,-4)     
\mt(3,-4)\te(--$a$--)\mt(43,-4)\te(--$b$--)\EndDraw&=&     
\Draw\ps(0.8mm)\la(3,3,43,3)\mt(23,-4)\te(--{\tiny A}--)\EndDraw~     
{\fh_a\fh_b\ov \fh^2}     
+\Draw\ps(0.8mm)\la(3,3,43,3)\mt(23,-4)\te(--{\tiny B}--)\EndDraw~     
\biggl[\d_{ab}-{\fh_a\fh_b\ov \fh^2}\biggr]\nn\\     
&=&{i\hbar\ov (1+\d Z)k^2-\d m^2-(\l+\d\l)\fh^2/2}     
{\fh_a\fh_b\ov \fh^2}\nn\\     
&&+{i\hbar\ov (1+\d Z)k^2-\d m^2-(\l+\d\l)\fh^2/6}     
\biggl[\d_{ab}-{\fh_a\fh_b\ov \fh^2}\biggr]\;,\nn\\     
\Draw\pra(-1,0,1,15)       
\RotatedAxes(120,210)\pra(-1,0,1,15)\EndRotatedAxes       
\RotatedAxes(240,330)\pra(-1,0,1,15)\EndRotatedAxes     
\mt(-14,-13)\te(--$a$--)\mt(17,-13)\te(--$b$--)\mt(0, 21)\te(--$c$--)     
\EndDraw     
&=&{1\ov 3}(\d_{ab}\fh_c+\d_{bc}\fh_a+\d_{ca}\fh_b)       
\biggl[-{i(\l+\d\l)\ov \hbar}\biggr]\;,\nn\\     
\Draw\RotatedAxes(45,135)\pra(-13,1,17,3)\EndRotatedAxes       
\RotatedAxes(45,135)\pra(1,-13,3,17)\EndRotatedAxes     
\mt(-14,-14)\te(--$a$--)\mt(16,21)\te(--$d$--)       
\mt(-14, 19)\te(--$c$--)\mt(16,-14)\te(--$b$--)\EndDraw       
&=&{1\ov 3}(\d_{ab}\d_{cd}+\d_{ac}\d_{bd}+\d_{ad}\d_{bc})     
\biggl[-{i(\l+\d\l)\ov \hbar}\biggr]\;,\label{jf}     
\eea     
where the propagators without group indices $a$ and $b$,      
but with small capital letters A and B are given, as self-evident in the      
above expression, by    
\bea     
\Draw\ps(0.8mm)\la(3,3,43,3)\mt(23,-4)\te(--{\tiny A}--)\EndDraw~     
&=&{i\hbar\ov (1+\d Z)k^2-\d m^2-(\l+\d\l)\fh^2/2}\;,\nn\\     
\Draw\ps(0.8mm)\la(3,3,43,3)\mt(23,-4)\te(--{\tiny B}--)\EndDraw~     
&=&{i\hbar\ov (1+\d Z)k^2-\d m^2-(\l+\d\l)\fh^2/6}\;.\label{pab}     
\eea     
The rules for vertices without group indices can also be defined:     
\bea     
\Draw\RotatedAxes(45,135)\pra(-13,1,17,3)\EndRotatedAxes       
\RotatedAxes(45,135)\pra(1,-13,3,17)\EndRotatedAxes\EndDraw     
=-{i(\l+\d\l)\ov \hbar}\;,~~~     
\Draw\pra(-1,0,1,15)       
\RotatedAxes(120,210)\pra(-1,0,1,15)\EndRotatedAxes       
\RotatedAxes(240,330)\pra(-1,0,1,15)\EndRotatedAxes\EndDraw       
=-{i(\l+\d\l)\sqrt{\fh^2}\ov \hbar}\;.\label{vert}     
\eea     
The propagators and vertices without the group indices,     
Eqs.~(\ref{pab}) and (\ref{vert}), will appear     
in the final formal expression when the group indices are contracted out.     
Without introducing any new loop-expansion parameter, which is eventually  
set to be unity, we will use $\hbar$ as a loop-counting parameter \cite{nb}.  
This is the reason why we have kept all the traces of $\hbar$'s in the  
Feynman rules, Eqs.~(\ref{jf}), (\ref{pab}), and (\ref{vert}), in spite of  
our employment of the usual ``God-given'' units, $\hbar=c=1$.  
In addition to the above Feynman rules, Eq.~(\ref{jf}), which are     
used in constructing two- and higher-loop vacuum diagrams, we need another      
set of rules solely for one-loop vacuum diagrams which are dealt with      
separately in Jackiw's prescription and are essentially      
the same as those of Coleman and Weinberg \cite{cw} from the outset:     
\bea     
\Draw\la(3,4.3,43,4.3)\la(3,1.7,43,1.7)\mt(23,-4)\te(--{\tiny A}--)       
\EndDraw&=&-\ln\biggl(1       
-{(\l+\d\l)\fh^2/2\ov (1+\d Z)k^2-\d m^2}\biggr)\;,\nn\\      
\Draw\la(3,4.3,43,4.3)\la(3,1.7,43,1.7)\mt(23,-4)\te(--{\tiny B}--)       
\EndDraw&=&-\ln\biggl(1       
-{(\l+\d\l)\fh^2/6\ov (1+\d Z)k^2-\d m^2}\biggr)\;.  \label{one}       
\eea    
   
Using the rules, Eqs.~(\ref{jf}) and (\ref{one}), and including the terms    
of zero-loop order, we arrive at the formal expression of the effective     
potential up to the three-loop order (see Fig.~1 to Fig.~9):     
\bea     
V_{\rm eff}(\fh_a)&=&\biggl[{\d m^2\ov 2}\fh^2     
+{\l+\d\l\ov 4!}\fh^4\biggr]+\Bigl[\mbox{Diag.~1}\Bigr]     
+\Bigl[\mbox{Diag.~2}\Bigr]+\Bigl[\mbox{Diag.~3}\Bigr]\nn\\  
&+&\Bigl[\mbox{Diag.~4}\Bigr]+\Bigl[\mbox{Diag.~5}\Bigr] 
+\Bigl[\mbox{Diag.~6}\Bigr]+\Bigl[\mbox{Diag.~7}\Bigr] 
+\Bigl[\mbox{Diag.~8}\Bigr]+\Bigl[\mbox{Diag.~9}\Bigr]\;.\label{onn}     
\eea  
We have revealed both symmetry number factor and group factor (the 
second factor when two factors appear simultaneously)      
explicitly in front of each diagram in Fig.~1 to Fig.~9.   
For the purpose of renormalization we first expand the counterterm  
constants in power series, beginning with order $\hbar$:  
\beas     
\d m^2&=&\hbar\d m_1^2+\hbar^2\d m_2^2+\cdots\;,\nn\\  
\d\l&=&\hbar\d \l_1+\hbar^2\d\l_2+\cdots\;,\nn\\  
\d Z&=&\hbar\d Z_1+\hbar^2\d Z_2+\cdots\;.       
\eeas     
As is well known in our original $\f^4$ theory of Eq.~(\ref{lg}) 
$\d Z_1$ vanishes \cite{cw,jk}. Thus   
the field renormalization counterterm appears in the    
three-loop calculation for the first time.  
In the first square-bracket term on the right-hand side of Eq.~(\ref{onn}), 
we need $\hbar^0$-, $\hbar^1$-, $\hbar^2$-, and $\hbar^3$-order  
terms in its expansion. In calculating [Diag.~1],  [Diag.~2], and [Diag.~3]  
terms, the careful $\hbar$ expansions are needed: $\hbar^1$-, $\hbar^2$-,  
and $\hbar^3$-order terms are needed from [Diag.~1] and $\hbar^2$-  
and $\hbar^3$-order terms from [Diag.~2] and [Diag.~3]. In the Appendix B,
the details of the counterterm integrals are described.
In the remaining terms, [Diag.~4] to [Diag.~9], the lowest order in $\hbar$    
is $\hbar^3$, which is already of the desired order. Thus with a simple  
replacement of every $\s^2$ in Eqs.~(\ref{3t}) and (\ref{3g})  with 
$\l\fh^2/2$ we readily evaluate the three-loop diagrams in Fig.~4 to  
Fig.~9. 
  
In what follows we will use the following notation for the effective potential 
up to the $L$-loop order: 
\beas 
V_{\rm eff}^{[L]}(\fh_a)=\sum_{i=0}^L \hbar^i  
V_{\rm eff}^{(i)}(\fh_a)\;, 
\eeas 
and use $\~{N}$ for $N-1$. 

\subsection{Up to Two-Loop Order}

The zero-loop part of the effective potential is given as 
\beas 
V_{\rm eff}^{(0)}(\fh_a)={\l\ov 4!}\fh^4\;. 
\eeas 
We first remove all the $\e$ poles in the subsequent contributions to the 
effective potential, $V_{\rm eff}^{(i)}(\fh_a)$ ($i$=1,2,3).  
The one-loop part of the effective potential is readily obtained as 
\bea 
V_{\rm eff}^{(1)}(\fh_a)&=&{\d m_1^2\ov 2}\fh^2+{\d \l_1\ov 4!} 
\fh^4 -{\l^2\fh^4\ov (4\p)^2\e}\biggl[{1\ov 8}+{\~{N}\ov 72}\biggr]     
+{\l^2\fh^4\ov (4\p)^2}\biggl[-{3\ov 32}+{\g\ov 16}+{1\ov 16}   
\ln\biggl({\l\fh^2/2\ov 4\p M^2}\biggr)\nn\\ 
&+&\~{N}\biggl\{-{1\ov 96}+{\g\ov 144}+{1\ov 144}   
\ln\biggl({\l\fh^2/6\ov 4\p M^2}\biggr)\biggr\}\biggr]\;.\label{v1p} 
\eea 
The $\e$ poles in this equation are readily cancelled out by choosing 
the counterterm constants $\d m_1^2$ and $\d\l_1$ as follows: 
\beas     
\d m_1^2&=&{\l\ov (4\p)^2}a_1\;,\nn\\     
\d\l_1&=&{\l^2\ov (4\p)^2}\biggl[{1\ov \e}\biggl(3+
{\~{N}\ov 3}\biggr)+b_1\biggr]\;, 
\eeas 
where $a_1$ and $b_1$ are unspecified but finite constants at this stage.  
One may put $a_1$ (and $a_2$, $a_3$ below) to be zero from the beginning 
because the theory is massless. In our dimensional regularization scheme 
the pole part of $\d m_1^2$ vanishes, but this is not true in the cutoff  
regularization method. 
 
The two-loop part of the effective potential is obtained as  
\bea 
V_{\rm eff}^{(2)}(\fh_a)&=&{\d m_2^2\ov 2}\fh^2+{\d \l_2\ov 4!}\fh^4     
-{a_1\l^2\fh^2\ov (4\p)^4\e}\biggl[{1\ov 2}+{\~{N}\ov 6}\biggr] -{\l^3\fh^4\ov     
(4\p)^4\e^2}\biggl[{3\ov 8}+{\~{N}\ov 12}+{\~{N}^2\ov 216}\biggr] \nn\\     
&+&{\l^3\fh^4\ov(4\p)^4\e}\biggl[{1\ov 8}+{5\~{N}\ov 216}\biggr]      
-{b_1\l^3\fh^4\ov (4\p)^4\e}\biggl[{1\ov 4}+{\~{N}\ov 36}\biggr] \nn\\     
&+&{a_1\l^2\fh^2\ov (4\p)^4}\biggl[-{1\ov 4}+{\g\ov 4}+{1\ov 4}\ln\biggl(     
{\l\fh^2/2\ov 4\p M^2}\biggr)+\~{N}\biggl\{-{1\ov 12}+{\g\ov 12}+{1\ov 12}     
\ln\biggl({\l\fh^2/6\ov 4\p M^2}\biggr)\biggr\}\biggr] \nn\\     
&+&{b_1\l^3\fh^4\ov (4\p)^4}\biggl[-{1\ov 8}+{\g\ov 8}+{1\ov 8}\ln\biggl(     
{\l\fh^2/2\ov 4\p M^2}\biggr)+\~{N}\biggl\{-{1\ov 72}+{\g\ov 72}+{1\ov 72}     
\ln\biggl({\l\fh^2/6\ov 4\p M^2}\biggr)\biggr\}\biggr] \nn\\     
&+&{\l^3\fh^4\ov(4\p)^4}\biggl[{11\ov 32}+{A\ov 8}-{5\ov 16}\g 
+{3\ov 32}\g^2+\biggl(    
-{5\ov 16}+{3\ov 16}\g\biggr)\ln\biggl({\l\fh^2/2\ov 4\p M^2}\biggr)
+{3\ov 32}\ln^2\biggl({\l\fh^2/2\ov 4\p M^2}\biggr)\nn\\    
&+&\~{N}\biggl\{{29\ov 432}+{A\ov 108} 
-{7\ov 108}\g+{\g^2\ov 48}+{13\ov 432}\ln 3-{\g\ov 48}\ln 3\nn\\     
&&~~~~~~~+\biggl(-{7\ov 108}
+{\g\ov 24}\biggr)\ln\biggl({\l\fh^2/2\ov 4\p M^2}\biggr)     
+{1\ov 48}\ln\biggl({\l\fh^2/2\ov 4\p M^2}\biggr)     
\ln\biggl({\l\fh^2/6\ov 4\p M^2}\biggr)\biggr\}\nn\\     
&+&\~{N}^2\biggl\{{1\ov 864}-{\g\ov 432}+{\g^2\ov 864}+ 
\biggl(-{1\ov 432}
+{\g\ov 432}\biggr)\ln\biggl({\l\fh^2/6\ov 4\p M^2}\biggr)     
+{1\ov 864}\ln^2\biggl({\l\fh^2/6\ov 4\p M^2}\biggr)\biggr\}\biggr]    
\;.\label{v2p}    
\eea 
Notice that the so-called ``dangerous'' pole terms such    
as $(\fh^l/\e)\ln [\l\fh^2/(4\p M^2)]$, $(l=0,2,4)$   
in the above equation, which    
cannot be removed by terms of counterterm constants [$\d m^2 \fh^2/2$    
and $\d\l\fh^4/(4!)$], have been completely     
cancelled out among each other. This serves as a strong check of the     
correctness of the calculation at this stage. 
The counterterm constants $\d m_2^2$ and  
$\d\l_2$ are determined as 
\beas     
\d m_2^2&=&{\l^2\ov (4\p)^4}\biggl[{a_1\ov \e}
\biggl(1+{\~{N}\ov 3}\biggr)+a_2\biggr]\;,\nn\\     
\d\l_2&=&{\l^3\ov (4\p)^4}\biggl[{1\ov \e^2}
\biggl(9+2\~{N}+{\~{N}^2\ov 9}\biggr)     
-{1 \ov \e}\biggl(3+{5\~{N}\ov 9}\biggr) 
+{b_1\ov \e}\biggl(6+{2\~{N}\ov 3}\biggr)+b_2\biggr]\;, 
\eeas    
where $a_2$ and $b_2$ are also unspecified but finite constants.  

In the minimal subtraction scheme (MS scheme), the finite parts of the  
counterterm constants ($a_1$, $b_1$, $a_2$, and $b_2$) 
are taken to be zero. In our massless theory, however, we encounter the 
infrared singularity in the defining condition for a coupling constant. 
To avoid this difficulty we follow Coleman and Weinberg \cite{cw} 
and require  
\bea   
{d^2 V_{\rm eff}\ov d \fh^2}\bigg |_{\fh=0}=0\;,~~~~   
{d^4 V_{\rm eff}\ov d \fh^4}\bigg |_{\fh=M}=\l\;,\label{rc}   
\eea   
where $M$ is a renormalization scale.    
Then constants $a_1$, $b_1$, $a_2$, and $b_2$ are determined as   
\bea   
a_1&=&a_2=0\;,\nn\\     
b_1&=&-4-{3\ov 2}\g-{3\ov 2}\ln\biggl({\l/2\ov    
4\p}\biggr)-\~{N}\biggl\{{4\ov 9}+{\g\ov 6}+{1\ov 6}\ln\biggl( 
{\l/6\ov 4\p}\biggr)\biggr\}\;,\nn\\    
b_2&=&{139\ov 4}-3A+15\g+{9\ov 4}\g^2   
+\biggl(15+{9\ov 2}\g\biggr)\ln\biggl({\l/2\ov 4\p}\biggr)   
+{9\ov 4}\ln^2\biggl({\l/2\ov 4\p}\biggr)\nn\\   
&+&\~{N}\biggl\{{202\ov 27}-{2\ov 9}A+{29\ov 9}\g+{\g^2\ov 2}   
-{14\ov 9}\ln 3-{\g\ov 2}\ln 3\nn\\   
&+&\biggl({29\ov 9}+\g-{1\ov 2}\ln 3\biggr)   
\ln\biggl({\l/2\ov 4\p}\biggr)+{1\ov 2}   
\ln^2\biggl({\l/2\ov 4\p}\biggr)\biggr\}\nn\\   
&+&\~{N}^2\biggl\{{113\ov 324}+{4\ov 27}\g+{\g^2\ov 36}   
+\biggl({4\ov 27}+{\g\ov 18}\biggr)\ln\biggl({\l/6\ov 4\p}\biggr)   
+{1\ov 36}\ln^2\biggl({\l/6\ov 4\p}\biggr)\biggr\}\;. \label{a12b12} 
\eea  
Putting these constants into the finite parts of $V_{\rm eff}^{(1)}$ and  
$V_{\rm eff}^{(2)}$, we eventually obtain 
the finite effective potential, $V_{\rm eff}^{[2]}$,    
which satisfies the renormalization conditions, Eq.~(\ref{rc}):  
\bea    
V_{\rm eff}^{[2]}(\fh_a)&=& 
\Biggl[{\l\ov 4!}\fh^4\Biggr] 
+{\hbar\l^2\fh^4\ov (4\p)^2}\Biggl[-{25\ov 96}+{1\ov 16}   
\ln\biggl({\fh^2\ov M^2}\biggr)   
+\~{N}\biggl\{-{25\ov 864}+{1\ov 144}\ln\biggl({\fh^2\ov M^2}\biggr)   
\biggr\}\Biggr]\nn\\   
&+&{\hbar^2\l^3\fh^4\ov (4\p)^4}\Biggl[{55\ov 24}-{13\ov 16}   
\ln\biggl({\fh^2\ov M^2}\biggr)+{3\ov 32}   
\ln^2\biggl({\fh^2\ov M^2}\biggr)+\~{N}\biggl\{{635\ov 1296}-{19\ov 108}   
\ln\biggl({\fh^2\ov M^2}\biggr)\nn\\ 
&+&{1\ov 48}\ln^2\biggl({\fh^2\ov M^2}\biggr)\biggr\} 
+\~{N}^2\biggl\{{85\ov 3888}-{11\ov 1296}\ln\biggl({\fh^2\ov M^2}\biggr)   
+{1\ov 864}\ln^2\biggl({\fh^2\ov M^2}\biggr)\biggr\}\Biggr]\;.
\eea

\subsection{Three-Loop Order}

Having illustrated our strategy, we continue with three-loop structure
of the model, which is appreciably more complicated than that of previous
subsection II.A. In the $\hbar^3$-order calculation we need another 
counterterm constant    
$\d Z_2$ in addition to the counterterm constants, $\d m_1^2$, $\d \l_1$, 
$\d m_2^2$, $\d \l_2$, ($\d m_3^2$, and $\d \l_3$) \cite{jm}. 
Instead of determining $\d Z_2$ from the renormalization of $\cz$ in     
Eq.~(\ref{loc}) \cite{iim,gkf}, we can obtain it from the 
renormalization of $\~{\G}^{(2)}_{a,b}(p,-p)$:   
\beas   
\d Z_2={\l^2\ov (4\p)^4}\biggl[{1\ov \e}\biggl(-{1\ov 12}-{\~{N}\ov 36}\biggr)
+c_2\biggr]\;,   
~~~~(c_2~\, \mbox{a finite constant})\;. 
\eeas    
Using the integrals in Appendices A and B, we have the rather long
expression for the three-loop part of the effective potential. 
Thus we will separate it into two parts, the counterterm plus pole part and
the finite part. The counterterm plus pole part is calculated as
\bea 
[V_{\rm eff}^{(3)}]_{\rm {ct+pl}}&=&   
{\d m^2_3\ov 2}\fh^2+{\d\l_3\ov 4!}\fh^4   
-{a_1\l^3\fh^2\ov (4\p)^6\e^2}\biggl[1+{7\~{N}\ov 18}   
+{\~{N}^2\ov 18}\biggr]\nn\\   
&+&{\l^3\fh^2\ov (4\p)^6\e}\biggl[a_1\biggl\{{1\ov 4}
+\~{N}\biggl({5\ov 36}-{\g\ov 9}+{1\ov 9}\ln 2\biggr)\biggr\}   
-a_1b_1\biggl\{{1\ov 2}+{\~{N}\ov 6}\biggr\}
-a_2\biggl\{{1\ov 2}+{\~{N}\ov 6}\biggr\}\biggr]\nn\\   
&-&{\l^4\fh^4\ov (4\p)^6\e^3}\biggl[{9\ov 8}+{3\~{N}\ov 8}   
+{\~{N}^2\ov 24}+{\~{N}^3\ov 648}\biggr]\nn\\
&+&{\l^4\fh^4\ov (4\p)^6\e^2}\biggl[{31\ov 36}+{41\~{N}\ov 162}   
+{17\~{N}^2\ov 972}\biggr]
-{b_1\l^4\fh^4\ov (4\p)^6\e^2}\biggl[{9\ov 8}+{\~{N}\ov 4}   
+{\~{N}^2\ov 72}\biggr]\nn\\
&-&{\l^4\fh^4\ov (4\p)^6\e}\biggl[{49\ov 192}+{\z(3)\ov 6}+\~{N}\biggl\{   
{253\ov 3888}+{5\ov 162}\z(3)\biggr\} 
+{35\~{N}^2\ov 15552}\biggr] \nn\\
&+&{\l^4\fh^4\ov (4\p)^6\e}\biggl[b_1\biggl\{{3\ov 8}+{5\~{N}\ov 72}\biggr\}  
-b_1^2\biggl\{{1\ov 8}+{\~{N}\ov 72}\biggr\}   
+c_2\biggl\{{1\ov 4}+{\~{N}\ov 36}\biggr\}
-b_2\biggl\{{1\ov 4}+{\~{N}\ov 36}\biggr\}\biggr]\;.\nn 
\eea
Notice that the complete cancellation of all  
dangerous pole terms [$(\fh^l/\e^m)\ln^n (\l\fh^2/(4\p M^2))$, $l=0,2,4$;
$(m,n)=(1,1)$, $(1,2)$, $(2,1)$] has taken place here too. 
Remaining harmless pole  
terms are eliminated by choosing the counterterm constants as follows:   
\bea   
\d m_3^2&=&{\l^3\ov (4\p)^6}\biggl[{a_1\ov \e^2}\biggl\{2+{7\~{N}\ov 9}   
+{\~{N}^2\ov 9}\biggr\}-{a_1\ov \e}\biggl\{{1\ov 2}   
+\~{N}\biggl({5\ov 18}-{2\ov 9}\g+{2\ov 9}\ln 2\biggr)\biggr\}\nn\\   
&+&{a_1b_1\ov \e}\biggl\{1+{\~{N}\ov 3}\biggr\}   
+{a_2\ov \e}\biggl\{1+{\~{N}\ov 3}\biggr\}+a_3\biggr]\nn\\   
\d \l_3&=&{\l^4\fh^4\ov (4\p)^6}\biggl[{1\ov \e^3}\biggl\{27+9\~{N}+\~{N}^2   
+{\~{N}^3\ov 27}\biggr\}\nn\\
&-&{1\ov \e^2}\biggl\{{62\ov 3}+{164\~{N}\ov 27}   
+{34\~{N}^2\ov 81}\biggr\}
+{b_1\ov \e^2}\biggl\{27+6\~{N}+{\~{N}^2\ov 3}\biggr\}\nn\\   
&+&{1\ov \e}\biggl\{{49\ov 8}+4\z(3)+\~{N}\biggl(   
{253\ov 162}+{20\ov 27}\z(3)\biggr)
+{35\~{N}^2\ov 648}\biggr\}\nn\\
&-&{b_1\ov \e}\biggl\{9+{5\~{N}\ov 3}\biggr\}   
+{b_1^2\ov \e}\biggl\{3+{\~{N}\ov 3}\biggr\}   
+{b_2\ov \e}\biggl\{6+{2\~{N}\ov 3}\biggr\}   
-{c_2\ov \e}\biggl\{6+{2\~{N}\ov 3}\biggr\}+b_3\biggr]\;,\label{dl3}
\eea   
where $a_3$ and $b_3$ are arbitrary but finite constants as before.
when taken in the MS scheme, $\d l_3$ in Eq.~\ref{dl3} compeletly 
coincides with that of Ref.~\cite{sft} which was obtained from the
renormalization of the $\G^{(2)}(p^2)$ and $\G^{(4)}(p^2)$. This fact
shows another check for the correctness of our calculation.    

The finite part is given as follows:  
\bea
[V_{\rm eff}^{(3)}]_{\rm fn}&=&{a_1^2\l^2\ov (4\p)^6}
\biggl[{1\ov 4}+{\~{N}\ov 4}\biggr]   
\ln\biggl({\l\fh^2\ov 4\p M^2}\biggr)\nn\\
&+&{\l^3\fh^2\ov (4\p)^6}\biggl[a_1\biggl\{   
-{3\ov 8}+{\g\ov 2}   
-{1\ov 2}\ln 2+\~{N}\biggl(-{2\ov 9}+{13\ov 36}\g   
-{13\ov 36}\ln 2-{1\ov 6}\ln 3\biggr)\nn\\   
&+&\~{N}^2\biggl(-{1\ov 72}+{\g\ov 36}-{1\ov 36}\ln 6\biggr)\biggr\} 
+a_2\biggl\{{1\ov 4}+{\~{N}\ov 12}\biggr\}
+a_1b_1\biggl\{{1\ov 4}+{\~{N}\ov 12}\biggr\}
\biggr]\ln\biggl({\l\fh^2\ov 4\p M^2}\biggr)\nn\\
&+&{a_1\l^3\fh^2\ov (4\p)^6}\biggl[{1\ov 4}+{13\~{N}\ov 72}+   
{\~{N}^2\ov 72}\biggr]\ln^2\biggl({\l\fh^2\ov 4\p M^2}\biggr)\nn\\
&+&{\l^4\fh^4\ov (4\p)^6}\biggl[
{701\ov 384}+{9\ov 16}A-{143\ov 96}\g   
+{27\ov 64}\g^2+{\z(3)\ov 4}+{143\ov 96}\ln 2-{27\ov 32}\g\ln 2 
+{27\ov 64}\ln^2 2\nn\\   
&+&\~{N}\biggl({2741\ov 5184}+{5\ov 48}A-{199\ov 432}\g   
+{9\ov 64}\g^2+{5\ov 108}\z(3)+{199\ov 432}\ln 2-{9\ov 32}\g\ln 2
+{9\ov 64}\ln^2 2\nn\\
&+&{125\ov 864}\ln 3-{3\ov 32}\g\ln 3   
+{3\ov 32}\ln 2\ln 3\biggr)
+\~{N}^2\biggl({365\ov 10368}+{A\ov 216}-{103\ov 2592}\g   
+{\g^2\ov 64}\nn\\
&+&{103\ov 2592}\ln 2-{\g\ov 32}\ln 2+{1\ov 64}\ln^2 2   
+{67\ov 2592}\ln 3-{\g\ov 48}\ln 3+{1\ov 48}\ln 2\ln 3+{1\ov 192}\ln^2 3   
\biggr)\nn\\   
&+&\~{N}^3\biggl({1\ov 5184}-{\g\ov 1296}+{\g^2\ov 1728}   
+{1\ov 1296}\ln 6-{\g\ov 864}\ln 6+{1\ov 1728}\ln^2 6\biggr)\nn\\
&+&b_1\biggl\{-{3\ov 4}+{9\ov 16}\g   
-{9\ov 16}\ln 2+\~{N}\biggl(-{11\ov 72}+{\g\ov 8}   
-{1\ov 8}\ln 2-{1\ov 16}\ln 3\biggr)\nn\\   
&+&\~{N}^2\biggl(-{1\ov 216}+{\g\ov 144}-{1\ov 144}\ln 6\biggr)\biggr\}
+b_1^2\biggl\{{1\ov 16}+{\~{N}\ov 144}\biggr\}
+b_2\biggl\{{1\ov 8}+{\~{N}\ov 72}\biggr\}\nn\\ 
&-&c_2\biggl\{{1\ov 8}+{\~{N}\ov 72}\biggr\}\biggr]   
\ln\biggl({\l\fh^2\ov 4\p M^2}\biggr)\nn\\
&+&{\l^4\fh^4\ov (4\p)^6}\biggl[
-{143\ov 192}+{27\ov 64}\g   
-{27\ov 64}\ln 2+\~{N}\biggl(-{199\ov 864}+{9\ov 64}\g   
-{9\ov 64}\ln 2-{3\ov 64}\ln 3\biggr)\nn\\   
&+&\~{N}^2\biggl(-{103\ov 5184}+{\g\ov 64}-{1\ov 64}\ln 2   
-{1\ov 96}\ln 3\biggr)+\~{N}^3\biggl(-{1\ov 2592}+{\g\ov 1728}
-{1\ov 1728}\ln 6\biggr)\nn\\
&+&b_1\biggl\{{9\ov 32}+{\~{N}\ov 16}+   
{\~{N}^2\ov 288}\biggr\}\biggr]\ln^2\biggl({\l\fh^2\ov 4\p M^2}   
\biggr)\nn\\
&+& {\l^4\fh^4\ov (4\p)^6}\biggl[{9\ov 64}+{3\~{N}\ov 64}   
+{\~{N}^2\ov 192}+   
{\~{N}^3\ov 5184}\biggr]\ln^3\biggl({\l\fh^2\ov 4\p M^2}\biggr)   
+\cdots\;,\label{v3p} 
\eea 
where three dots ($\cdots$) denote finite terms which can be absorbed   
into the counterterms. 

If we choose the boundary condition  
$d\~{\G}^{(2)}(p^2)/d p^2|_{p^2=\l M^2/2}=1$, the constant $c_2$,  
which is used in obtaining $b_3$, is determined as   
\bea   
c_2={\l^2\ov (4\p)^4}\biggl({1\ov 12}+{\~{N}\ov 36}\biggr)\biggl\{   
-{9\ov 4}+\g+\ln\biggl({\l/2\ov 4\p}\biggr)\biggr\}\;. \label{c2}  
\eea  
From the renormalization conditions Eq.~(\ref{rc}) we obtain
\bea
a_3&=&0\;,\nn\\
b_3&=&-{27035\ov 48}-{75\ov 4}A-25\z(3)
+\biggl(-{1957\ov 8}-{9\ov 2}A-6\z(3)\biggr)\ln\biggl({\l\ov 4\p}\biggr)\nn\\
&&-{359\ov 8}\ln^2\biggl({\l\ov 4\p}\biggr)
-{27\ov 8}\ln^3\biggl({\l\ov 4\p}\biggr)\nn\\
&&+\~{N}\biggl\{-{228725\ov 1296}-{125\ov 36}A-{125\ov 27}\z(3)
+{175\ov 216}\ln 3\nn\\
&&+\biggl(-{16769\ov 216}-{5\ov 6}A-{10\ov 9}\z(3) 
+{7\ov 36}\ln 3\biggr)\ln\biggl({\l\ov 4\p}\biggr)    
-{523\ov 36}\ln^2\biggl({\l\ov 4\p}\biggr)
-{9\ov 8}\ln^3\biggl({\l\ov 4\p}\biggr)
\biggr\}\nn\\
&&+\~{N}^2\biggl\{-{62105\ov 3888}-{25\ov 162}A+{175\ov 648}\ln 3
+\biggl(-{4775\ov 648}-{A\ov 27}+{7\ov 108}\ln 3\biggr)
\ln\biggl({\l\ov 4\p}\biggr)\nn\\
&&-{319\ov 216}\ln^2\biggl({\l\ov 4\p}\biggr)
-{1\ov 8}\ln^3\biggl({\l\ov 4\p}\biggr)\biggr\}\nn\\ 
&&+\~{N}^3\biggl\{-{4655\ov 11664}-{395\ov 1944}\ln\biggl({\l\ov 4\p}\biggr)
-{5\ov 108}\ln^2\biggl({\l\ov 4\p}\biggr)
-{1\ov 216}\ln^3\biggl({\l\ov 4\p}\biggr)\biggr\}\;.\label{a3b3}
\eea
Putting the constants in Eqs.~(\ref{a12b12}), (\ref{c2}), and (\ref{a3b3})
into Eq.~(\ref{v3p}) we finally arrive at the three-loop part of the 
effective potential:
\bea 
V_{\rm eff}^{(3)}(\fh_a)&=& 
{\hbar^3\l^4\fh^4\ov (4\p)^6}\Biggl[   
-{27035\ov 1152}-{25\ov 32}A-{25\ov 24}\z(3) 
+\~{N}\biggl(-{228725\ov 31104}-{125\ov 864}A
-{125\ov 648}\z(3)+{175\ov 5184}\ln 3\biggr)\nn\\
&&+\~{N}^2\biggl(-{62105\ov 93312}-{25\ov 3888}A
+{175\ov 15552}\ln 3\biggr)- 
{4655\ov 279936}\~{N}^3+\biggl\{{1957\ov 192}+{3\ov 16}A+{\z(3)\ov 4}\nn\\ 
&&+\~{N}\biggl({16769\ov 5184}
+{5\ov 144}A+{5\ov 108}\z(3)
-{7\ov 864}\ln 3\biggr)+\~{N}^2\biggl({4775\ov 15552}+{A\ov 648}
-{7\ov 2592}\ln 3\biggr)\nn\\ 
&&+{395\ov 46656}\~{N}^3\biggr\}\ln\biggl({\fh^2\ov M^2}\biggr)
+\biggl\{-{359\ov 192}-{523\ov 864}\~{N}-{319\ov 5184}\~{N}^2-{5\ov 2592}   
\~{N}^3\biggr\}\ln^2\biggl({\fh^2\ov M^2}\biggr)\nn\\   
&&+\biggl\{{9\ov 64}+{3\ov 64}\~{N}+{\~{N}^2\ov 192}+{\~{N}^3\ov 5184}   
\biggr\}\ln^3\biggl({\fh^2\ov M^2}\biggr)\Biggr]\;.\label{3p}  
\eea  
\section{Concluding Remarks}      
 
In this paper, we have proceeded one loop further in the effective potential  
calculation of the massless $O(N)$ $\f^4$ theory, that is, we have calculated 
the three-loop  effective potential of the massless $O(N)$ $\f^4$ theory in  
four dimensions for the first time. 
For an easy check of the cancellation of all the 
dangerous pole terms among themselves in each loop order, 
we have given the $\e$-expanded loop integrals in full detail.   
The $\hbar^3$-order part of Eq.~(\ref{3p}) is our main result. 

We see that the order $\hbar$- and $\hbar^2$- terms completely   
agree with previous calculations \cite{jk} in which a different 
regularization   
scheme (cutoff regularization) was employed. In the cutoff regularization,  
the loop integrations in four dimensions are much more difficult in the   
three-loop diagrams with the overlap divergence. 

We expect the present calculations will serve as a useful reference
for a development of the three-loop effective potential of more realistic  
models containing gauge fields such as the scalar QED. Especially in relation 
to this, the Coleman-Weinberg mechanism is interesting, which appears in many
contexts, from displacive phase transitions in solids to the thermodynamics 
of the early universe.

\acknowledgments   
This work was supported in part by Ministry of Education, Project number               
BSRI-97-2442 and one of the authors (J.~-M. C.) was also supported in           
part by the Postdoctoral Fellowship of Kyung Hee University.

\appendix  
\section{Loop Integration Formulas}       
\setcounter{equation}{0} 
In this Appendix A, the momenta appearing in the formulas are all    
(Wick-rotated) Euclidean ones and the abbreviated integration measure is   
defined as    
\beas    
\int_k=M^{4-n}\int{d^n k\ov (2\p)^n}\;,    
\eeas    
where $n=4-\e$ is the space-time dimension in the framework of dimensional   
regularization \cite{tv} and $M$ is an arbitrary constant with mass  
dimension, usually taken as the renormalization scale.    
One-loop integrations are quite elementary. For the two-loop  
integrations one may refer to Ref.~\cite{2lp}. The genuine three-loop 
integrals
which cannot be factorized into the lower-loop ones are quite involved. 
Here we simply quote the results for them. Details of the computation 
can be found in Ref.~\cite{jmjm}.

\ni \underline{One-loop integrals}:  
\bea       
S_0&\equiv&\int_k\ln\biggl(1+{\xi^2\ov k^2+\s^2}\biggr)       
=-{(\xi^2+\s^2)^2\ov (4\p)^2}  
\nn\\   &\times&  
\biggl({\xi^2+\s^2\ov 4\p M^2}\biggr)^{\!\!\!-\e/2}\G\biggl({\e\ov 2}-2\biggr)  
+\,\xi\mbox{-independent term}\;,\nn\\   
S_1&\equiv&\int_k{1\ov k^2+\s^2}       
={\s^2\ov (4\p)^2}\biggl({\s^2\ov 4\p M^2}       
\biggr)^{\!\!\!-\e/2}\G\biggl({\e\ov 2}-1\biggr)\;,\nn\\       
S_2&\equiv&\int_k{1\ov k^2+\s^2/3}       
={\s^2\ov 3(4\p)^2}\biggl({\s^2/3\ov 4\p M^2}       
\biggr)^{\!\!\!-\e/2}\G\biggl({\e\ov 2}-1\biggr)\;,\nn\\       
S_3&\equiv&       
\int_k{1\ov (k^2+\s^2)^2}={1\ov (4\p)^2}\biggl({\s^2\ov 4\p M^2}       
\biggr)^{\!\!\!-\e/2}\G\biggl({\e\ov 2}\biggr)\;,\nn\\       
S_4&\equiv&       
\int_k{1\ov (k^2+\s^2/3)^2}={1\ov (4\p)^2}\biggl({\s^2/3\ov 4\p M^2}       
\biggr)^{\!\!\!-\e/2}\G\biggl({\e\ov 2}\biggr)\;.       
\eea     
\ni \underline{Two-loop integrals}	:       
\bea       
W_1&\equiv&\int_{kp}{1\ov (p^2+\s^2)[(p+k)^2+\s^2]}       
={\s^4\ov (4\p)^4}\biggl({\s^2\ov 4\p M^2}       
\biggr)^{\!\!\!-\e}\G^2\biggl({\e\ov 2}-1\biggr)\;,\nn\\       
W_2&\equiv&\int_{kp}{1\ov (p^2+\s^2/3)       
[(p+k)^2+\s^2/3]}={\s^4\ov 9(4\p)^4}\biggl({\s^2/3\ov 4\p M^2}       
\biggr)^{\!\!\!-\e}\G^2\biggl({\e\ov 2}-1\biggr)\;,\nn\\       
W_3&\equiv&\int_{kp}{1\ov (p^2+\s^2)       
[(p+k)^2+\s^2/3]}={\s^4\ov 3(4\p)^4}\biggl({\s^2\ov 4\p M^2}       
\biggr)^{\!\!\!-\e/2}\biggl({\s^2/3\ov 4\p M^2}\biggr)^{\!\!\!-\e/2}       
\G^2\biggl({\e\ov 2}-1\biggr)\;,\nn\\       
W_4&\equiv&\int_{kp}{1\ov (k^2+\s^2)       
(p^2+\s^2)[(p+k)^2+\s^2]}\nn\\      
&=&{\s^2\ov (4\p)^4}\biggl({\s^2\ov 4\p M^2}\biggr)^{\!\!\!-\e}       
{\G^2(1+\e/2)\ov (1-\e)(1-\e/2)}\biggl[-{6\ov \e^2}-3A+O(\e) 
\biggr]\;,\nn\\       
W_5&\equiv&     
\int_{kp}{1\ov (k^2+\s^2)(p^2+\s^2/3)[(p+k)^2+\s^2/3]}\nn\\       
&=&{\s^2\ov 3(4\p)^4}\biggl({\s^2/3\ov        
4\p M^2}\biggr)^{\!\!\!{-\e}}       
{\G^2(1+\e/2)\ov (1-\e)(1-\e/2)}\biggl[-{10\ov \e^2}+{6\ov \e}\ln 3 
-{3\ov 2}       
\ln^2 3-B+O(\e)\biggr]\;,\nn\\       
W_6&\equiv&     
\int_{kp}{1\ov (k^2+\s^2)^2(p^2+\s^2)[(p+k)^2+\s^2]}\nn\\       
&=&{1\ov (4\p)^4}\biggl({\s^2\ov 4\p M^2}       
\biggr)^{\!\!\!-\e}\biggl[{2\ov \e^2}+{1\ov \e}\biggl\{1-2\g\biggr\}+       
{1\ov 2}-\g+\g^2+{\p^2\ov 12}+A+O(\e)\biggr]\;,\nn\\       
W_7&\equiv&     
\int_{kp}{1\ov (k^2+\s^2/3)^2(p^2+\s^2/3)[(k+p)^2+\s^2]}\nn\\       
&=&{1\ov (4\p)^4}\biggl({\s^2/3\ov 4\p M^2}       
\biggr)^{\!\!\!-\e}\biggl[{2\ov \e^2}+{1\ov \e}\biggl\{1-2\g\biggr\}+       
{1\ov 2}-\g+\g^2+{\p^2\ov 12}+B+O(\e)\biggr]\;,\nn\\       
W_8&\equiv&     
\int_{kp}{1\ov (k^2+\s^2)^2(p^2+{\s^2\ov 3})[(k+p)^2+{\s^2\ov 3}]}\nn\\       
&=&{1\ov (4\p)^4}\biggl({\s^2\ov 4\p M^2}       
\biggr)^{\!\!\!-\e}\biggl[{2\ov \e^2}+{1\ov \e}\biggl\{1-2\g\biggr\}+       
{1\ov 2}-\g+\g^2+{\p^2\ov 12}+C+O(\e)\biggr]\;.\label{www}     
\eea       
In the above equation, $\g$ is the usual Euler constant,    
$\g=0.5772156649\cdots$, and numerical values of the constants,    
$A$, $B$, and $C$ in Eq.~(\ref{www}) are       
\beas    
A=f(1,1)=-1.1719536193\cdots\;,\nn\\      
B=f(1,3)=-2.3439072387\cdots\;,\nn\\      
C=f\Bigl({1\ov 3},{1\ov 3}\Bigr)=0.1778279325\cdots\;,      
\eeas     
where     
\beas     
&& f(a,b)\equiv\int_0^1dx\biggl[\int_0^{1-z}dy\biggl(-{\ln(1-y)\ov y} 
\biggr)       
-{z\ln z\ov 1-z}\biggr]\;,~~~z={ax+b(1-x)\ov x(1-x)}\;.     
\eeas          
These constants $A$, $B$, and $C$ can be analytically integrated \cite{kt}.
The results are expressed in terms of Clausen function:
\beas
A={B\ov 2}=-{3\ov 2}C-{3\ov 4}\ln^2 3=
-{2\ov \sqrt{3}}{\rm Cl}_2\Bigl({\p\ov 3}\Bigr)\;,
\eeas
where
\beas 
{\rm Cl}_2(\tt)=\int_0^\tt\ln[2\sin(\tt'/2)]d\tt'\;. 
\eeas

\ni \underline{Trivial three-loop integrals}:       
\bea       
H_1&\equiv&       
\int_k{1\ov (k^2+\s^2)^2}\biggl\{\int_p{1\ov p^2+\s^2}\biggr\}^{\!2}       
=S_1^2S_3\;,\nn\\       
H_2&\equiv&       
\int_k{1\ov (k^2+\s^2)^2}\int_p{1\ov p^2+\s^2}\int_q{1\ov q^2 
+\s^2/3}=S_1S_2S_3\;,\nn\\       
H_3&\equiv&       
\int_k{1\ov (k^2+\s^2/3)^2}\biggl\{\int_p{1\ov p^2+\s^2} 
\biggr\}^{\!2}=S_1^2S_4\;,\nn\\       
H_4&\equiv&       
\int_k{1\ov (k^2+\s^2)^2}\biggl\{\int_p{1\ov p^2+\s^2/3} 
\biggr\}^{\!2}       
=S_2^2S_3\;,\nn\\       
H_5&\equiv&\int_k{1\ov (k^2+\s^2/3)^2}     
\int_p{1\ov p^2+\s^2}\int_q{1\ov q^2+\s^2/3}       
=S_1S_2S_4\;,\nn\\       
H_6&\equiv&\int_k{1\ov (k^2+\s^2/3)^2}     
\biggl\{\int_p{1\ov p^2+\s^2/3}\biggr\}^{\!2}       
=S_2^2S_4\;,\nn\\       
I_1&\equiv&\int_k{1\ov k^2+\s^2}     
\int_{p,\,q}{1\ov (p^2+\s^2)^2(q^2+\s^2)[(p+q)^2+\s^2]}       
=S_1W_6\;,\nn\\       
I_2&\equiv&\int_k{1\ov k^2+\s^2/3}     
\int_{p,\,q}{1\ov (p^2+\s^2)^2(q^2+\s^2)[(p+q)^2+\s^2]}       
=S_2W_6\;,\nn\\       
I_3&\equiv&\int_k{1\ov k^2+\s^2}     
\int_{p,\,q}{1\ov (p^2+\s^2/3)^2(q^2+\s^2/3)[(p+q)^2+\s^2]}       
=S_1W_7\;,\nn\\       
I_4&\equiv&       
\int_k{1\ov k^2+\s^2/3}\int_{p,\,q}     
{1\ov (p^2+\s^2/3)^2(q^2+\s^2/3)[(p+q)^2+\s^2]}       
=S_2W_7\;,\nn\\       
I_5&\equiv&\int_k{1\ov k^2+\s^2}     
\int_{p,\,q}{1\ov (p^2+\s^2)^2(q^2+\s^2/3)[(p+q)^2+\s^2/3]}       
=S_1W_8\;,\nn\\       
I_6&\equiv&\int_k{1\ov k^2+\s^2/3}     
\int_{p,\,q}{1\ov (p^2+\s^2)^2(q^2+\s^2/3)[(p+q)^2+\s^2/3]}       
=S_2W_8\;.  \label{3t}   
\eea     
     
\ni \underline{Genuine three-loop integrals}:    
\bea     
J_1&\equiv&       
\int_k\biggl\{\int_p{1\ov (p^2+\s^2)[(p+k)^2+\s^2]}\biggr\}^{\!2} \nn\\      
&&=\Omega_2\biggl[{16\ov \e^3}+{1\ov \e^2}\biggl\{{92\ov 3}           
-24\g\biggr\}                 
+{1\ov \e}\biggl\{35-46\g+18\g^2+\p^2\biggr\}\biggr]\;,\nn\\       
J_2&\equiv&       
\int_{kpq}{1\ov (p^2+\s^2)[(p+k)^2+\s^2](q^2+\s^2/3)[(q+k)^2     
+\s^2/3]}\nn\\       
&=&\Omega_2               
\biggl[{176\ov 27\e^3}+{1\ov \e^2}\biggl\{{332\ov 27}-{88\ov 9}\g+               
{28\ov 9}\ln3\biggr\}            
\nn\\    &&~~~~~~            
+{1\ov \e}\biggl\{{365\ov 27}-{166\ov 9}\g+{22\ov 3}\g^2               
+{11\ov 27}\p^2+{55\ov 9}\ln 3-{14\ov 3}\g\ln 3+\ln^2 3\biggr\}           
\biggr]\;,\nn\\ 
J_3&\equiv&       
\int_k\biggl\{\int_p{1\ov (p^2+\s^2/3)     
[(p+k)^2+\s^2/3]}\biggr\}^{\!2}     
=\Omega_2               
\biggl[{16\ov 9\e^3}+{1\ov \e^2}\biggl\{{92\ov 27}-{8\ov 3}\g+{8\ov 3}                 
\ln 3\biggr\}            
\nn\\   &&~~~~~~            
+{1\ov \e}\biggl\{{35\ov 9}-{46\ov 9}\g+2\g^2+{\p^2\ov 9}                 
+{46\ov 9}\ln 3-4\g\ln 3+2\ln^2 3\biggr\}\biggr]\;,\nn\\  
K_1&\equiv&       
\int_k{1\ov k^2+\s^2}     
\biggl\{\int_p{1\ov (p^2+\s^2)[(p+k)^2+\s^2]}\biggr\}^{\!2} \nn\\       
&=&\Omega_1\biggl[-{8\ov \e^3}+{1\ov \e^2}\biggl\{-{68\ov 3}           
+12\g\biggr\}                 
+{1\ov \e}\biggl\{-{134\ov 3}-12A+34\g-9\g^2-{\p^2\ov 2}\biggr\}           
\biggr]\;,\nn\\       
K_2&\equiv&       
\int_{kpq}{1\ov (k^2+\s^2)(p^2+\s^2)[(p+k)^2+\s^2](q^2+\s^2/3)       
[(q+k)^2+\s^2/3]}\nn\\       
&=&\Omega_1\biggl[-{56\ov 9\e^3}+{1\ov \e^2}\biggl\{-{52\ov 3}           
+{28\ov 3}\g               
-{4\ov 3}\ln 3\biggr\}            
\nn\\      &&~~~~~~            
+{1\ov \e}\biggl\{-{302\ov 9}-6A-{2\ov 3}B+26\g-7\g^2-{7\ov 18}\p^2               
-4\ln 3+2\g\ln 3\biggr\}\biggr]\;,\nn\\        
K_3&\equiv&       
\int_k{1\ov k^2+\s^2/3}       
\biggl\{\int_p{1\ov (p^2+\s^2)[(p+k)^2+\s^2/3]}\biggr\}^{\!2}\nn\\     
&=&\Omega_1               
\biggl[-{40\ov 9\e^3}+{1\ov \e^2}\biggl\{-{116\ov 9}+{20\ov 3}\g           
-{8\ov 3}\ln 3               
\biggr\}\nn\\               
&&~~~~~~+{1\ov \e}\biggl\{-26-{4\ov 3}B+{58\ov 3}\g-5\g^2               
-{5\ov 18}\p^2+4\g\ln 3-{22\ov 3}\ln 3\biggr\}               
\biggr]\;,\nn\\   
K_4&\equiv&\int_k{1\ov k^2+\s^2}       
\biggl\{\int_p{1\ov (p^2+\s^2/3)[(p+k)^2+\s^2/3]} 
\biggr\}^{\!2}\nn\\       
&=&\Omega_1\biggl[-{40\ov 9\e^3}+{1\ov \e^2}\biggl\{-12+{20\ov 3}\g               
-{8\ov 3}\ln 3\biggr\}            
\nn\\   &&~~~~~~            
+{1\ov \e}\biggl\{-{202\ov 9}-{4\ov 3}B+18\g-5\g^2-{5\ov 18}\p^2-8\ln 3                 
+4\g\ln 3\biggr\}\biggr]\;,\nn\\ 
L_1&\equiv&       
\int_k{1\ov (k^2+\s^2)^2}     
\biggl\{\int_p{1\ov (p^2+\s^2)[(p+k)^2+\s^2]}\biggr\}^{\!2}\nn\\       
&=&\Omega_0\biggl[{8\ov 3\e^3}+{1\ov \e^2}\biggl\{{8\ov 3}-4\g\biggr\}                 
+{1\ov \e}\biggl\{{4\ov 3 }+4A-4\g+3\g^2+{\p^2\ov 6}\biggr\}           
\biggr]\;,\nn\\       
L_2&\equiv&       
\int_{kpq}{1\ov (k^2+\s^2)^2(p^2+\s^2)[(p+k)^2+\s^2](q^2 
+\s^2/3)       
[(q+k)^2+\s^2/3]}\nn\\       
&=&\Omega_0\biggl[{8\ov 3\e^3}+{1\ov \e^2}\biggl\{{8\ov 3}-4\g\biggr\}                 
+{1\ov \e}\biggl\{{4\ov 3 }+2A+2C-4\g+3\g^2+{\p^2\ov 6}\biggr\}           
\biggr]\;,\nn\\     
L_3&\equiv&\int_k{1\ov (k^2+\s^2/3)^2}       
\biggl\{\int_p{1\ov (p^2+\s^2)[(p+k)^2+\s^2/3]}\biggr\}^{\!2}\nn\\     
&=&\Omega_0               
\biggl[{8\ov 3\e^3}+{1\ov \e^2}\biggl\{{8\ov 3}-4\g+4\ln 3\biggr\}\nn\\            
&&~~~~~~            
+{1\ov \e}\biggl\{{4\ov 3 }+4B-4\g+3\g^2+{\p^2\ov 6}+4\ln 3-6\g\ln 3               
+3\ln^2 3\biggr\}\biggr]\;,\nn\\     
L_4&\equiv&\int_k{1\ov (k^2+\s^2)^2}       
\biggl\{\int_p{1\ov (p^2+\s^2/3)[(p+k)^2+\s^2/3]} 
\biggr\}^{\!2}\nn\\         
&=&\Omega_0\biggl[{8\ov 3\e^3}+{1\ov \e^2}\biggl\{{8\ov 3}-4\g\biggr\}                 
+{1\ov \e}\biggl\{{4\ov 3}+4C-4\g+3\g^2+{\p^2\ov 6}\biggr\}\biggr]\;,\nn\\    
M_1&\equiv&\int_{kpq}{1\ov (k^2+\s^2)(p^2+\s^2)(q^2+\s^2)       
[(k-p)^2+\s^2][(p-q)^2+\s^2][(q-k)^2+\s^2]} \nn\\       
&=&\Omega_0\biggl[{4\z(3)\ov \e}\biggr]\;,\nn\\  
M_2&\equiv&     
\int_{kpq}{1\ov (k^2+\s^2/3)(p^2+\s^2/3)(q^2+\s^2/3)       
[(k-p)^2+\s^2][(p-q)^2+\s^2][(q-k)^2+\s^2]}\nn\\      
&=&\Omega_0\biggl[{4\z(3)\ov \e}\biggr]\;,\nn\\         
M_3&\equiv&     
\int_{kpq}{1\ov (k^2+\s^2)(p^2+\s^2/3)(q^2+\s^2/3)       
[(k-p)^2+\s^2/3][(p-q)^2+\s^2][(q-k)^2+\s^2/3]}\nn\\      
&=&\Omega_0\biggl[{4\z(3)\ov \e}\biggr]\;. \label{3g}    
\eea  
In the above equation the overall multiplying factors are     
\beas     
\Omega_0={1\ov (4\p)^6}\biggl({\s^2\ov 4\p M^2} 
\biggr)^{\!\!\!{-3\e/2}}\;,~~~     
\Omega_1={\s^2\ov (4\p)^6}\biggl({\s^2\ov 4\p M^2} 
\biggr)^{\!\!\!{-3\e/2}}\;,   
~~~     
\Omega_2={\s^4\ov (4\p)^6}\biggl({\s^2\ov 4\p M^2}\biggr)^{\!\!\!{-3\e/2}}\;.    
\eeas    
\section{Details of the Connterterm Integrals}
\setcounter{equation}{0} 
Diagrams Fig.~1 to Fig.~3 contain various counterterm integrals in our 
perturbation theory. In this Appendix B they are shown explicitly:

\bea        
{\rm Diag.~1a}&=&-{\hbar\ov 2(4\p)^2}\biggl({\d m^2+        
(\l+\d\l)\fh^2/2\ov 1+\d Z}\biggr)^{\!\!2}\biggl({\d m^2+(\l+\d\l)        
\fh^2/2\ov 4\p M^2(1+\d Z)}\biggr)^{\!\!\!-\e/2}\G\biggl(-2+ 
{\e\ov 2}\biggr)\nn\\        
&=&\hbar\biggl[-{\l^2\fh^4\ov 8(4\p)^2}        
\biggl({\l\fh^2/2\ov 4\p M^2}\biggr)^{\!\!\!-\e/2}        
\G\biggl(-2+{\e\ov 2}\biggr)\biggr]\nn\\        
&+&\hbar^2\biggl[-{\l\fh^2\ov 2(4\p)^2}\biggl(\d m_1^2 
+{\d\l_1\fh^2\ov 2}\biggr)        
\biggl(1-{\e\ov 4}\biggr)\biggl({\l\fh^2/2\ov 4\p M^2} 
\biggr)^{\!\!\!-\e/2}\G\biggl(-2+{\e\ov 2}\biggr)\biggr]\nn\\        
&+&\hbar^3\biggl[-{1\ov 2(4\p)^2}\biggl\{        
\biggl(\d m_1^2+{\d\l_1\fh^2\ov 2}\biggr)^{\!\!2}\biggl(1-{3\e\ov 4}        
+{\e^2\ov 8}\biggr)\nn\\        
&+&\l\fh^2\biggl(\d m_2^2+{\d\l_2\fh^2\ov 2}\biggr)        
\biggl(1-{\e\ov 4}\biggr)+\l^2\fh^4\d Z_2\biggl(-{1\ov 2}+{\e\ov 8} 
\biggr)\biggr\}\biggl({\l\fh^2/2\ov 4\p M^2}\biggr)^{\!\!\!-\e/2}        
\G\biggl(-2+{\e\ov 2}\biggr)\biggr]\;,\nn\\        
\mbox{Diag.~1b}&=&-{\hbar(N-1)\ov 2(4\p)^2}\biggl({\d m^2+        
(\l+\d\l)\fh^2/6\ov 1+\d Z}\biggr)^{\!\!2}\biggl({\d m^2+(\l+\d\l)        
\fh^2/6\ov 4\p M^2(1+\d Z)}\biggr)^{\!\!\!-\e/2}\G\biggl(-2 
+{\e\ov 2}        
\biggr)\nn\\        
&=&\hbar\biggl[-{(N-1)\l^2\fh^4\ov 72(4\p)^2}        
\biggl({\l\fh^2/6\ov 4\p M^2}\biggr)^{\!\!\!-\e/2}        
\G\biggl(-2+{\e\ov 2}\biggr)\biggr]\nn\\        
&+&\hbar^2\biggl[-{(N-1)\l\fh^2\ov 6(4\p)^2}\biggl(\d m_1^2+        
{\d\l_1\fh^2\ov 6}\biggr)        
\biggl(1-{\e\ov 4}\biggr)\biggl({\l\fh^2/6\ov 4\p M^2} 
\biggr)^{\!\!\!-\e/2}\G\biggl(-2+{\e\ov 2}\biggr)\biggr]\nn\\        
&+&\hbar^3\biggl[-{N-1\ov 2(4\p)^2}\biggl\{        
\biggl(\d m_1^2+{\d\l_1\fh^2\ov 6}\biggr)^{\!\!2}\biggl(1-{3\e\ov 4}        
+{\e^2\ov 8}\biggr)\nn\\        
&+&{\l\fh^2\ov 3}\biggl(\d m_2^2+{\d\l_2\fh^2\ov 6}\biggr)        
\biggl(1-{\e\ov 4}\biggr)+{\l^2\fh^4\d Z_2\ov 9}\biggl(-{1\ov 2}        
+{\e\ov 8}\biggr)\biggr\}\biggl({\l\fh^2/6\ov 4\p M^2} 
\biggr)^{\!\!\!-\e/2}\G\biggl(-2+{\e\ov 2}\biggr)\biggr]\;,\nn\\        
\mbox{Diag.~2a}&=&{\hbar^2(\l+\d\l)\ov 8(4\p)^4(1+\d Z)^2}        
\biggl({\d m^2+(\l+\d\l)\fh^2/2\ov 1+\d Z}\biggr)^{\!\!2}        
\biggl({\d m^2+(\l+\d\l)\fh^2/2\ov 4\p M^2(1+\d Z)} 
\biggr)^{\!\!\!-\e}\G^2\biggl(-1+{\e\ov 2}\biggr)\nn\\        
&=&\hbar^2\biggl[{\l^3\fh^4\ov 32(4\p)^4} 
\biggl({\l\fh^2/2\ov 4\p M^2}        
\biggr)^{\!\!\!-\e}\G^2\biggl(-1+{\e\ov 2}\biggr)\biggr]\nn\\        
&+&\hbar^3\biggl[{1\ov 8(4\p)^4}\biggl\{\l^2\fh^2\d m_1^2 
\biggl(1-{\e\ov 2}        
\biggr)+\l^2\fh^4\d\l_1\biggl({3\ov 4}-{\e\ov 4}\biggr)\biggr\}        
\biggl({\l\fh^2/2\ov 4\p M^2}\biggr)^{\!\!\!-\e}        
\G^2\biggl(-1+{\e\ov 2}\biggr)\biggr]\;,\nn\\        
\mbox{Diag.~2b}&=&{\hbar^2(N-1)(\l+\d\l)\ov 12(4\p)^4(1+\d Z)^2}        
\biggl({\d m^2+(\l+\d\l)\fh^2/2\ov 1+\d Z}\biggr)        
\biggl({\d m^2+(\l+\d\l)\fh^2/6\ov 1+\d Z}\biggr)\nn\\        
&&\times        
\biggl({\d m^2+(\l+\d\l)\fh^2/2\ov 4\p M^2(1+\d Z)} 
\biggr)^{\!\!\!-\e/2}        
\biggl({\d m^2+(\l+\d\l)\fh^2/6\ov 4\p M^2(1+\d Z)} 
\biggr)^{\!\!\!-\e/2}        
\G^2\biggl(-1+{\e\ov 2}\biggr)\nn\\        
&=&\hbar^2\biggl[{(N-1)\l^3\fh^4\ov 144(4\p)^4}        
\biggl({\l\fh^2/2\ov 4\p M^2}\biggr)^{\!\!\!-\e/2}        
\biggl({\l\fh^2/6\ov 4\p M^2}\biggr)^{\!\!\!-\e/2}        
\G^2\biggl(-1+{\e\ov 2}\biggr)\biggr]\nn\\        
&+&\hbar^3\biggl[{N-1\ov 12(4\p)^4}\biggl\{\l^2\fh^2\d m_1^2        
\biggl({2\ov 3}-{\e\ov 3}\biggr)+\l^2\fh^4\d\l_1\biggl({1\ov 4}        
-{\e\ov 12}\biggr)\biggr\}\nn\\        
&&\times        
\biggl({\l\fh^2/2\ov 4\p M^2}\biggr)^{\!\!\!-\e/2}        
\biggl({\l\fh^2/6\ov 4\p M^2}\biggr)^{\!\!\!-\e/2}        
\G^2\biggl(-1+{\e\ov 2}\biggr)\biggr]\;,\nn\\        
\mbox{Diag.~2c}&=&{\hbar^2(N^2-1)(\l+\d\l)\ov 24(4\p)^4(1+\d Z)^2}        
\biggl({\d m^2+(\l+\d\l)\fh^2/6\ov 1+\d Z}\biggr)^{\!\!2}        
\biggl({\d m^2+(\l+\d\l)\fh^2/6\ov 4\p M^2(1+\d Z)} 
\biggr)^{\!\!\!-\e}        
\G^2\biggl(-1+{\e\ov 2}\biggr)\nn\\        
&=&\hbar^2\biggl[{(N^2-1)\l^3\fh^4\ov 864(4\p)^4}        
\biggl({\l\fh^2/6\ov 4\p M^2}        
\biggr)^{\!\!\!-\e}\G^2\biggl(-1+{\e\ov 2}\biggr)\biggr]\nn\\        
&+&\hbar^3\biggl[{N^2-1\ov 24(4\p)^4}\biggl\{{\l^2\fh^2\d m_1^2\ov 3}        
\biggl(1-{\e\ov 2}\biggr)+{\l^2\fh^4\d\l_1\ov 9}\biggl({3\ov 4}        
-{\e\ov 4}\biggr)\biggr\}\biggl({\l\fh^2/6\ov 4\p M^2}  
\biggr)^{\!\!\!-\e}\G^2\biggl(-1+{\e\ov 2}\biggr)\biggr]\;,\nn\\        
\mbox{Diag.~3a}&=&-{\hbar^2(\l+\d\l)^2\fh^2\ov 12(4\p)^4(1+\d Z)^3}        
\biggl({\d m^2+(\l+\d\l)\fh^2/2\ov 1+\d Z}\biggr)\biggl(        
{\d m^2+(\l+\d\l)\fh^2/2\ov 4\p M^2(1+\d Z)} 
\biggr)^{\!\!\!-\e}\nn\\        
&&\times{\G^2(1+\e/2)\ov (1-\e)(1-\e/2)}\biggl(-{6\ov \e^2}-3A 
\biggr)\nn\\        
&=&\hbar^2\biggl[-{\l^3\fh^4\ov 24(4\p)^4} 
\biggl({\l\fh^2/2\ov 4\p M^2}        
\biggr)^{\!\!\!-\e}{\G^2(1+\e/2)\ov (1-\e)(1-\e/2)}\biggl(-{6\ov \e^2}        
-3A\biggr)\biggr]\nn\\        
&+&\hbar^3\biggl[-{1\ov 12(4\p)^4}\biggl\{\l^2\fh^2\d m_1^2(1-\e)        
+\l^2\fh^4\d\l_1\biggl({3\ov 2}-{\e\ov 2}\biggr)\biggr\}        
\biggl({\l\fh^2/2\ov 4\p M^2}\biggr)^{\!\!\!-\e}\nn\\        
&&\times        
{\G^2(1+\e/2)\ov (1-\e)(1-\e/2)}\biggl(-{6\ov \e^2}-3A\biggr) 
\biggr]\;,\nn\\        
\mbox{Diag.~3b}&=&-{\hbar^2(N-1)(\l+\d\l)^2\fh^2\ov 36(4\p)^4(1+\d Z)^3}        
\biggl({\d m^2+(\l+\d\l)\fh^2/6\ov 1+\d Z}\biggr)\biggl(        
{\d m^2+(\l+\d\l)\fh^2/6\ov 4\p M^2(1+\d Z)} 
\biggr)^{\!\!\!-\e}\nn\\        
&&\times{\G^2(1+\e/2)\ov (1-\e)(1-\e/2)}\biggl(-{10\ov \e^2}        
+{6\ov \e}\ln 3-{3\ov 2}\ln^2 3-B\biggr)\biggr]\nn\\        
&=&\hbar^2\biggl[-{(N-1)\l^3\fh^4\ov 216(4\p)^4}        
\biggl({\l\fh^2/6\ov 4\p M^2}        
\biggr)^{\!\!\!-\e}{\G^2(1+\e/2)\ov (1-\e)(1-\e/2)}\biggl(-{10\ov \e^2}        
+{6\ov \e}\ln 3-{3\ov 2}\ln^2 3-B\biggr)\biggr]\nn\\        
&+&\hbar^3\biggl[-{N-1\ov 36(4\p)^4}\biggl\{\l^2\fh^2\d m_1^2(1-\e)        
+\l^2\fh^4\d\l_1\biggl({1\ov 2}-{\e\ov 6}\biggr)\biggr\}        
\biggl({\l\fh^2/6\ov 4\p M^2}\biggr)^{\!\!\!-\e}\nn\\        
&&\times{\G^2(1+\e/2)\ov (1-\e)(1-\e/2)}\biggl(-{10\ov \e^2}        
+{6\ov \e}\ln 3-{3\ov 2}\ln^2 3-B\biggr)\biggr]\;.\nn     
\eea     


\begin{figure}[h] 
  {\unitlength1cm 
   \epsfig{file=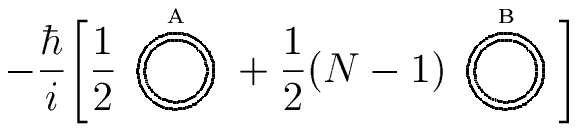, 
   bbllx=85pt,bblly=0pt,bburx=612pt,bbury=660pt, 
      rheight=2.5cm, rwidth=10cm,  clip=,angle=0} 
    }   
\caption{[Diag.~1]\,=\,[Diag.~1a]+\,[Diag.~1b].}   
\end{figure}

\begin{figure}[h] 
  {\unitlength1cm 
   \epsfig{file=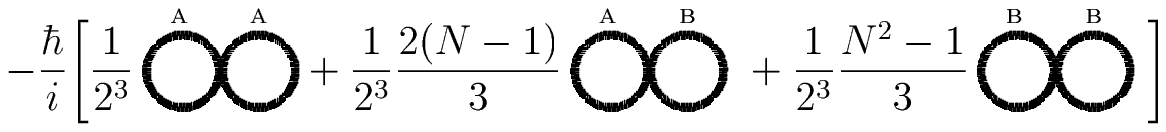, 
   bbllx=90pt,bblly=0pt,bburx=612pt,bbury=660pt, 
      rheight=2.5cm, rwidth=10cm, clip=,angle=0}   
    }   
\caption{[Diag.~2]\,=\,[Diag.~2a]\,+\,[Diag.~2b]\,+\,[Diag.~2c].} 
\end{figure}

\begin{figure}[h] 
  {\unitlength1cm 
   \epsfig{file=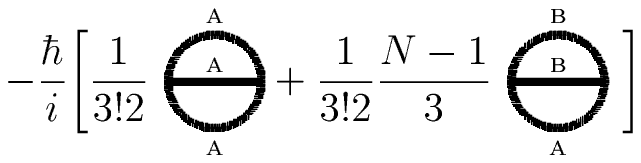, 
   bbllx=85pt,bblly=0pt,bburx=612pt,bbury=650pt, 
      rheight=2.5cm, rwidth=10cm,  clip=,angle=0} 
    }   
\caption{[Diag.~3]\,=\,[Diag.~3a]\,+\,[Diag.~3b].} 
\end{figure}

\begin{figure}[h] 
  {\unitlength1cm 
   \epsfig{file=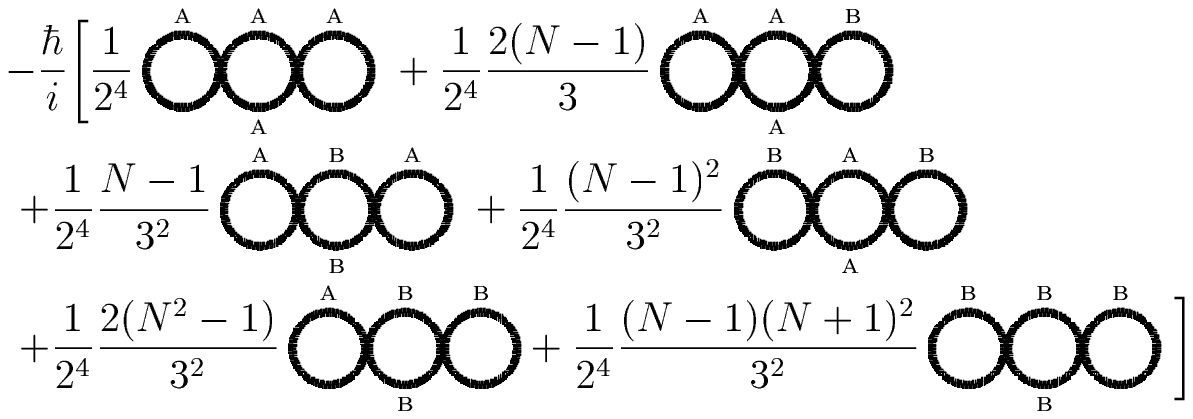, 
   bbllx=90pt,bblly=0pt,bburx=612pt,bbury=650pt, 
      rheight=5cm, rwidth=10cm,  clip=,angle=0} 
    }   
\caption{[Diag.~4]\,=\,[Diag.~4a]\,+\,[Diag.~4b]\,+\,[Diag.~4c]\, 
+\,[Diag.~4d]\,+\,[Diag.~4e]\,+\,[Diag.~4f].} 
\end{figure}

\begin{figure}[h] 
  {\unitlength1cm 
   \epsfig{file=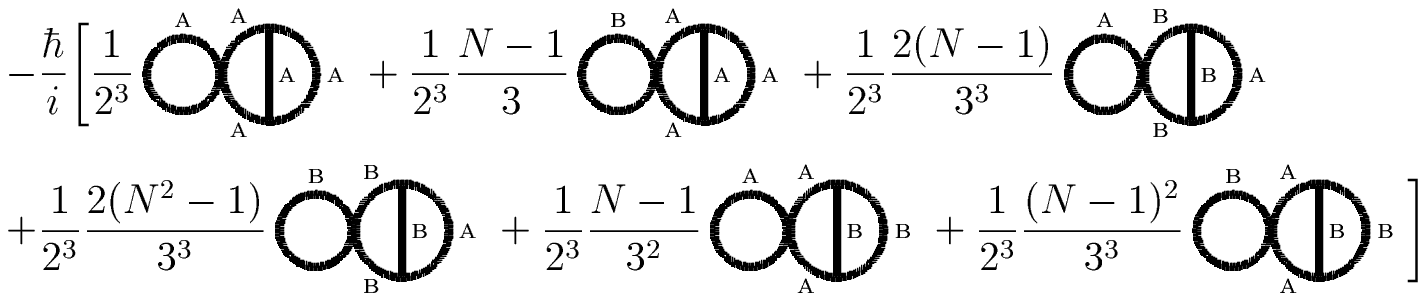, 
   bbllx=120pt,bblly=0pt,bburx=612pt,bbury=650pt, 
      rheight=4cm, rwidth=13cm,  clip=,angle=0} 
    }   
\caption{[Diag.~5]\,=\,[Diag.~5a]\,+\,[Diag.~5b]\,+\,[Diag.~5c]\, 
+[\,Diag.~5d]\,+\,[Diag.~5e]\,+\,[Diag.~5f].}   
\end{figure}

\begin{figure}[h] 
  {\unitlength1cm 
   \epsfig{file=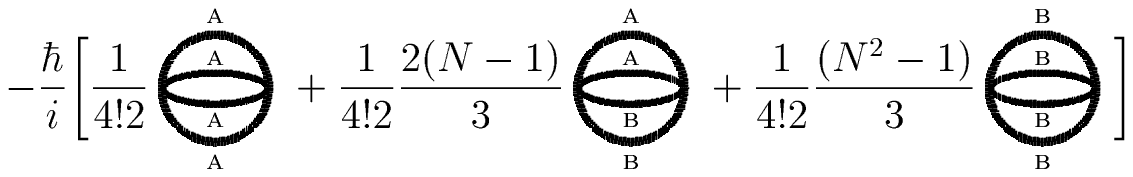, 
   bbllx=90pt,bblly=0pt,bburx=612pt,bbury=660pt, 
      rheight=3cm, rwidth=10cm, clip=,angle=0} 
    }   
\caption{[Diag.~6]\,=\,[Diag.~6a]\,+\,[Diag.~6b]\,+\,[Diag.~6c].}     
\end{figure}

\begin{figure}[h] 
  {\unitlength1cm 
   \epsfig{file=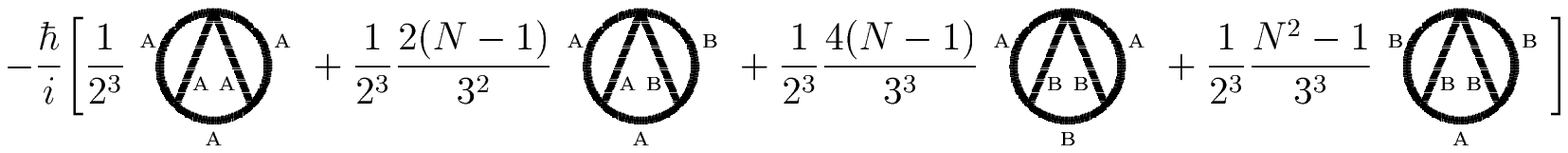, 
   bbllx=115pt,bblly=0pt,bburx=612pt,bbury=660pt, 
      rheight=3cm, rwidth=16cm,  
      clip=,angle=0} 
    }   
\caption{[Diag.~7]\,=\,[Diag.~7a]\,+\,[Diag.~7b]\,+\,[Diag.~7c]\, 
+\,[Diag.~7d].}  
\end{figure}

\begin{figure}[h] 
  {\unitlength1cm 
   \epsfig{file=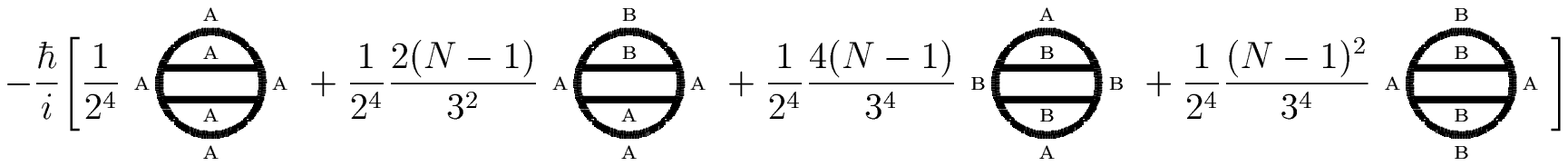, 
   bbllx=115pt,bblly=0pt,bburx=612pt,bbury=660pt, 
      rheight=3cm, rwidth=16cm,  
      clip=,angle=0} 
    }   
\caption{[Diag.~8]\,=\,[Diag.~8a]\,+\,[Diag.~8b]\,+\,[Diag.~8c]\, 
+\,[Diag.~8d].}   
\end{figure}

\begin{figure}[h] 
  {\unitlength1cm 
   \epsfig{file=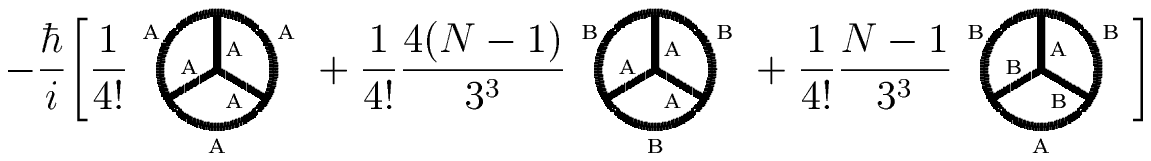, 
   bbllx=90pt,bblly=0pt,bburx=612pt,bbury=660pt, 
      rheight=3cm, rwidth=10cm,  
      clip=,angle=0} 
    }   
\caption{[Diag.~9]\,=\,[Diag.~9a]\,+\,[Diag.~9b]\,+\,[Diag.~9c].}     
\end{figure}

~\\
~\\

\end{document}